\documentstyle[12pt,a4]{article}

\begin{document}
                     
\bibliographystyle{plain}

\noindent
\centerline{{\large \bf LIFE EXTINCTIONS DUE TO}} 
\centerline{{\large \bf  NEUTRON STAR  MERGERS}} 

\medskip
\noindent
\centerline{{\bf Arnon Dar, Ari 
Laor, and Nir J. Shaviv}}

\medskip
\noindent
\centerline {Department of Physics and Space Research Institute}
\centerline {Israel 
Institute of Technology, Haifa 32000, Israel.}

\begin{abstract}

Cosmic ray bursts (CRBs) from mergers or accretion induced collapse of
neutron stars that hit an Earth-like planet closer than $\sim 1 ~kpc$ from
the explosion produce lethal fluxes of atmospheric muons at ground level,
underground and underwater. These CRBs also destroy the ozone layer and
radioactivate the environment.  The mean rate of such life devastating
CRBs is one in 100 million years ($Myr$), consistent with the observed 5
``great'' extinctions in the past $600~Myr$. Unlike the previously
suggested extraterrestrial extinction mechanisms the CRBs explain massive
life extinction on the ground, underground and underwater and the higher
survival levels of radiation resistant species and of terrain sheltered
species. More distant mergers can cause smaller extinctions.  Biological
mutations caused by ionizing radiation produced by the CRB may explain a
fast appearance of new species after mass extinctions. The CRB extinction
predicts detectable enrichment of rock layers which formed during the
extinction periods with cosmogenically produced radioactive nucleides such
as $^{129}$I, $^{146}$Sm, $^{205}$Pb with and $^{244}$Pu. Tracks of high
energy particles in rock layers on Earth and on the moon may also
contain records of intense CRBs.  An early warning of future extinctions
due to neutron star mergers can be obtained by identifying, mapping and
timing all the nearby binary neutron stars systems. A final warning of an
approaching CRB from a nearby neutron stars merger will be provided
by a gamma ray burst a few days before the arrival of the CRB.
\end{abstract}    
\clearpage  

\section{Introduction}

The early history of single-celled organisms during the Precambrian, 4560
to 570 million years  ($Myr$) ago, is poorly known. Since the end of the
Precambrian the diversity of both marine and continental life increased
exponentially. Analysis of fossil record of microbes, algae, fungi,
protists, plants and animals shows that this diversification was
interrupted by five major mass extinction ``events'' and some smaller
extinction peaks (1) . The ``big five'' mass extinctions occurred in the
Late Ordovician, Late Devonian Late Permian, Late Triassic and
end-Cretaceous and included both marine and continental life. The largest
extinction occurred about $251~Myr$ ago at the end of the Permian
period. The global species extinction ranged then between 80\% to 95\%,
much more than, for instance, the end-Ordovician extinction $439~Myr$ ago 
which eliminated 57\% of marine genera, or the
Cretaceous-Tertiary extinction $64~Myr$ ago which killed the
dinosaurs and claimed 47\% of existing genera (2). In spite of intensive
studies it is still not known what caused the mass extinctions, how quick
were they and whether they were subject to regional variations.  Many
extinction mechanisms have been proposed but no single mechanism seems to
provide a satisfactory explanation of both the marine and continental
extinction levels, the biological extinction patterns and the repetition
rate of the mass extinctions (1,2). These include astrophysical extinction
mechanisms, such as meteoritic impact that predicts the iridiun
anomaly which was found at the Cretaceous/Tertiary boundary (3) but has
not been found in all the other extinctions (4), supernova explosions (5) and
gamma ray bursts (6) which do not occur close enough at sufficiently high
rate to explain the observed rate of mass extinctions.
In this paper we propose that cosmic ray bursts (CRBs) from neutron
star (NS) mergers at distances closer than 1 kpc from planet Earth
produced lethal fluxes of atmospheric muons at ground level, 
underground and underwater, destroyed the ozone layer radioactivated
the atmosphere and the surface of the planet and polluted it with 
radioactive isotopes. Such CRBs can explain the 
large mass extinctions on planet Earth in the past $600~Myr.$ 
Unlike other extraterrestrial extinction mechanisms that have been
proposed before, CRBs are much more effective in extincting life and can
explain massive extinction underwater and underground, the higher
survival levels of radiation resistant species and terrain sheltered
species, and the observed rate of the
large mass extinctions on Earth.  More distant galactic mergers may have
caused some of the smaller extinctions on our planet. Biological mutations
induced by ionizing radiation could have caused a fast appearance of new
species after mass extinctions.

\section {CRBs From NS Mergers}

Three NS-NS binaries are presently known
within the galactic disc; PSRs B1913+16 (7) at a distance of $D\sim 7.3~
kpc$, B203+46 (8) at $D\sim 2.3~kpc$, 
and B1534+12 (9)  at $D\sim 0.5~kpc$,
and PSR 2127+11C (10) in the globular cluster M15 at $D\sim 10.6~kpc$.
Although unseen as radio
pulsars, the companion stars in these systems have been identified as
neutron stars from their mass which is inferred via measurements of the
relativistic periastron advance using standard pulse timing techniques (11).
These binary neutron stars loose energy by gravitational wave emission.
The continuous energy loss brings them closer and closer until they finally  
merge. Their merger times due to gravitational radiation have been
calculated (12) from the observed binary period, orbital 
eccentricity and
the masses of the pulsar and its companion. They are $3.01\times 10^8~yr$,
$t\gg t_H~$, $2.73\times 10^9~yr$ and $2.20\times 10^8~yr$, respectively,
where $t_H=1/H$ is the Hubble age of the Universe. It has been estimated
that there are $\sim 240$ potentially observable NS-NS binaries in the MW
disc and their merging rate is (13) $R_{MW}\sim 2-3\times 
(Myr)^{-1}$. 
Such NS-NS mergers release enormous amount of gravitational binding energy,
$\sim M_\odot c^2$, in few ms in the form of
gravitational waves, neutrinos and kinetic energy of ejected 
material (14).  The ejecta probably consists of high Z material 
because the crust of neutron stars consists of iron-like 
nuclei and because even if the ejecta is initially made of 
neutrons or dissociated nuclear matter, upon expansion nucleosynthesis  
transforms it quickly into the most bound, iron-like nuclei (15). 
The kinetic energy of the ejecta can be estimated 
from numerical simulations of
NS-NS mergers (16). Such estimates are highly
questionable since numerical simulations have difficulties 
in  reproducing even Type II supernovae explosions, let alone
the much more complicated NS-NS mergers. However, 
a rough estimate of this energy can be obtained directly from observations. 
Typically in SNeII, $10M_\odot$ are accelerated to a final velocity of 
(12) $v\sim 10^4 ~km~s^{-1}$, i.e., to a total final momentum $P\sim10
M_\odot v$. Since core collapse is not affected directly by the 
stellar envelope, we assume that a similar impulse, $\int Fdt\approx P$,
is imparted to the ejected mass in merger/AIC   collapse of NS (this
may not be the case if energy and momentum deposition of neutrinos
in the envelope is driving the explosion). 
If the ejected mass is much smaller than a solar mass, $\Delta M \ll
M_\odot$, it is accelerated to a highly relativistic velocity. Its kinetic
energy then is, $E_K \sim Pc\sim 5\times 10^{52}d\Omega~erg$ in a solid  
angle $d\Omega$, which is injected 
into the interstellar medium as a very powerful cosmic ray burst (CRB). 

\section{Indirect Evidence From GRBs}

It was shown (17) that highly relativistic ejecta from NS-NS mergers
can boost and beam star light in dense stellar regions into cosmological
gamma ray bursts (GRBs) very similar to those observed (18), provided
its kinetic energy is $\sim 5\times 10^{52}d\Omega~erg$ and
$\Gamma=E/mc^2\sim 1000$, where $d\Omega> 1/4\Gamma^2$ is the solid angle
of the beamed emission. In particular, such CRBs in dense stellar regions
explain remarkably well (17) the burst size, bimodal duration
distribution, complex light curves and spectral evolution of GRBs.
Moreover, the estimated rate of NS-NS mergers in the Universe (19) is
consistent with the observed rate (18) of GRBs and the
cosmic ray injection rate in the MW that is required to maintain a
constant cosmic ray flux in the MW: 

The escape rate of cosmic rays from the MW requires an average injection
rate of $ Q_{CR}\sim 5\times 10^{40}~erg~s^{-1}$, in high energy cosmic
rays above GeV, in order to maintain a constant energy density of cosmic
rays in the MW (20), where we used a total mass of gas $\sim 4.8\times
10^9M_\odot$ in the MW disc deduced from cosmic ray production of galactic
$\gamma$ rays. This injection rate is consistent with that due to NS-NS
mergers in the disc of the MW,
\begin{equation}
Q_{CR}^{NS} \sim R_{MW} E_K\sim 5\times10^{40}~erg~s^{-1}. 
\end{equation}

\section{Attenuation of CRBs}

The typical burst size and duration of GRBs suggest that the energy
release by NS-NS merger in high energy cosmic rays into a solid angle
$d\Omega > 1/4\Gamma^2$ in the direction of Earth is $\sim 5\times
10^{52}d\Omega~ erg$ and that $\Gamma=E/mc^2\sim 1000$ (17). Such CRBs
cannot be attenuated significantly by the column densities of radiation
and gas in the galactic disk. This is unlike the non relativistic ejecta
in supernova (SN) explosions, which is attenuated by the interstellar gas
over a distance of a few $pc$ by Coulomb collisions:  Moderately energetic
charged particles, other than electrons, lose energy in neutral
interstellar gas primarily by ionization. The mean rate of energy loss (or
stopping power) is given by the Bethe-Bloch formula 
\begin{equation}
-dE/dx\approx (4\pi Z^2 n_e \beta^{-2}e^4/m_e
c^2)[(1/2)ln(2m_ec^2\beta^2\gamma^2T_{max}/I^2)-\beta^2], 
\end{equation}
where $Ze$ is the charge of the energetic particle of mass $M_i$, velocity
$\beta c$ and total energy $E=\gamma M_ic^2$, $n_e$ is the number of
electrons per unit volume in the medium in atoms with ionization potential
$I$, and $T_{max}=2m_ec^2\beta^2\gamma^2 /(1+2\gamma m_e/M_i)$. Hence, for
SN ejecta with $Z\sim 1$ and $\beta=v/c\sim 10^{-2}$ in a neutral
interstellar medium with a typical density of $n_{H}\sim 1~cm^{-3}$ the
stopping distance of the ejecta due to Coulomb interactions is $x\sim
E/2(dE/dx)\sim 0.7~pc$. If the interstellar gas around the SN is ionized,
then $I$ has to be replaced by $e^2/R_D$ where $R_D=(kT/4\pi
e^2n_e(Z+1))^{1/2}$ is the Debye screening length. Thus for an ionized
interstellar gas with $kT\sim 1~eV$ the range is $x\sim E/2(dE/dx) \sim
6~pc$ if $v\sim 10000~km~s^{-1}$. In fact, the stopping of the ejecta by
Coulomb interactions is consistent with observations of SN remnants, like
SN 1006 (21), while the assumption that the ionized interstellar medium is
glued to the swept up magnetic field seems to be contradicted by some
recent observations (22). The stopping power of the interstellar medium
for highly relativistic iron-like nuclei, which dominate the CRB
composition, is due mainly to nuclear collisions and radiative energy losses
and not to ionization losses. The range of iron-like nuclei with $\Gamma\sim
1000$ in hydrogen is approximately $\sim 10^{25} cm $ in a typical
interstellar density of $n_{H}\sim 1~cm^{-3}$, while their range due only
to ionization energy loss is $\sim 10^{28}~ cm$. 

Although the galactic magnetic field, $H\sim 3-5~\times 10^{-6}~ Gauss$,
results in a Larmor radius of $r_{L}=\beta\Gamma m c/qH\sim 10^{14}~cm$
for protons with $\Gamma=1000$, it cannot deflect and disperse the CRB
particles from a NS-NS merger at distance smaller than $\sim 1~kpc$ from
the explosion. This is similar to the case of highly relativistic jets
from active galactic nuclei which reach distances of hundreds of kpc
without significant deflection.  Because of its high energy density, the
CRB sweeps up the interstellar magnetic field as long as its distance from
the explosion is smaller than $D\sim (6E_K/H^2)^{1/3}\sim 2~kpc$, where
its energy density becomes comparable to that of the swept up magnetic
field. (The energy densities of the galactic cosmic rays and the magnetic
field in the galactic disc are $\epsilon_{CR}^G\sim \epsilon_{H}^G\sim
10^{-12}~erg~cm^{-3}$, respectively). Thus,  CRBs expand ballistically,
like jets from active galactic nuclei, up to a distance $D\sim 1~kpc$. 

\section{Mass Extinction By CRBs} 

We assume that the ambient interstellar gas is not swept up with the CRB.
If it were, then the CRB would not expand beyond $\sim 10~pc$. In SN
explosions collective modes are invoked as the source of the coupling of
the ejecta to the interstellar medium, required in order to attenuate the
SN debris. As mentioned above, binary Coulomb interactions are sufficient
to produce the observed coupling in SN explosions, and coupling through
collective modes is not necessarily present. 

Unattenuated CRBs from NS-NS mergers can be devastating to life on 
nearby planets. At a distance of $1~kpc$ the duration of unattenuated  
CRB is \begin{equation} \delta t\sim
D/2c\Gamma^2\sim 1~day - 2~months 
\end{equation} 
for typical values of $\Gamma$ between 1000 and 100, respectively. The
time integrated energy flux of the CRB at $D\sim 1~kpc$ is, typically,
$3\times 10^{12}~GeV~cm^{-2}$. Thus, the energy deposition in the
atmosphere by the CRB is equivalent to the total energy deposition of
galactic cosmic rays in the atmosphere over $\sim 10^5~ yr$.  However, the
typical energy of the cosmic rays in the CRB is $\sim 1~TeV$ per nucleon,
compared with $\sim 1~GeV$ per nucleon for ordinary cosmic ray nuclei. 
Collisions of such CRB particles in the atmosphere
of a planet generate atmospheric cascades where a significant fraction
of the CRB  energy is converted into
``atmospheric muons'' through leptonic decay modes of the produced mesons. 
Most of these muons do not decay in the atmosphere because of their high 
energy,  unlike most of the 
atmospheric muons that are produced by ordinary 
cosmic rays . The average number of high energy muons produced by 
nucleons of primary
energy $E_p$, which do not decay in the atmosphere and reach sea level
with energy $>E_\mu$ at zenith angle $\theta<\pi/2$, is given
approximately by (23)
\begin{equation}
<N_\mu>\sim 
(0.0145E_\mu[TeV])(E_0/E_\mu)^{0.757}(1-E_\mu/E_0)^{5.25}/cos\theta~.
\end{equation}
Thus a CRB with energy of about $1~TeV$ per nucleon at a distance of
$1~kpc$ produces at sea level a flux of atmospheric muons of 
\begin{equation}
 I_\mu(>3~GeV)\sim 10^{12}~ cm^{-2}.          
\end{equation} 
Such muons deposit energy in matter via ionization. Their energy
deposition rate is (24) $-dE/dx\geq 2~MeV~g^{-1}cm^{-1}$.  The whole-body
lethal dose from penetrating ionizing radiation resulting in 50\%
mortality of human beings in 30 days (24) is $\leq 300~rad \sim 2\times
10^{10}/(dE/dx) \sim 10^{10}~cm^{-2}$ where $dE/dx$ is in rate is in
$MeV~g^{-1}cm^{-1}$ units.  The lethal dosages for other vertebrates can
be a few times larger while for insects they can be as much as a factor 20
larger.  Hence, a CRB at $D\sim 1~kpc$ which is not significantly
dispersed by the galactic magnetic field produces a highly lethal burst of
atmospheric muons. Because of muon penetration, the large muon flux is
lethal for most species even deep (hundreds of meters) underwater and
underground, if the cosmic rays arrive from well above the horizon. Thus,
unlike the other suggested extraterrestrial extinction mechanisms, a CRB
which produces a lethal burst of atmospheric muons can explain also
massive extinction deep underwater and why extinction was higher in
shallow waters. 

Although half of the planet is in 
the shade of the CRB, planet rotation exposes a larger fraction of the 
planet surface to the CRB. Additional effects increase the 
lethality of the CRB over the whole planet. They include: 
\noindent
(a) The pollution of the environment by radioactive nuclei, produced by
spallation of atmospheric and surface nuclei by shower particles. Using
the analytical methods of ref. 25, we estimate that for an Earth-like
atmosphere, the flux of energetic nucleons which reaches the surface is
also considerable,    
\begin{equation}    
    I_p(>100~MeV)\sim I_n (>100~MeV)\sim 10^{10}~cm^{-2}. 
    \end{equation}
Global winds spread radioactive gases in a relatively 
short time over the whole planet.

\noindent
(b) Depletion of stratospheric ozone by the reaction of ozone with nitric
oxide, generated by the cosmic ray produced electrons in the atmosphere
(massive destruction of stratospheric ozone has been observed during large
solar flares which produced energetic protons (26)). 

\noindent
(c) Extensive damage to the food chain by radioactive pollution and
massive extinction of vegetation and living organisms by ionizing
radiations (the lethal radiation dosages for trees and plants are slightly
higher than those for animals but still less than the flux given by
eq. 5 for all except the most resilient species). 

\section{Signatures of CRB Extinction}

{\bf The biological extinction pattern:} The biological extinction pattern
due to a CRB depends on the exposure and the vulnerability of the
different species to the primary and secondary effects of the CRB. The
exposure of the living organisms to the muon burst depends on the size and
duration of the CRB, on its direction relative to the rotation axis of
Earth (Earth shadowing), on the local sheltering provided by terrain
(canyons, mountain shades) and by underwater and underground habitats, and on
the risk sensing/assessment and mobility of the various species. The 
lethality of the
CRB depends as well on the vulnerability of the various living species and
vegetation to the primary ionizing radiation, to the drastic changes in
the environment (e.g., radioactive pollution and destruction of the ozone
layer) and to the massive damage and radioactive poisoning of the food
chain. Although the exact biological signature may be quite complicated,
and somewhat obscured in fossil records (due to poor or limited sampling,
deterioration of the rocks with time and dating and interpretation
uncertainties because of bioturbational smearing) it may show the 
general pattern expected from a CRB
extinction. Indeed, a first examination of the fossil records suggest that
there is a clear correlation between the extinction pattern of
different species, their vulnerability to ionizing radiation and the
sheltering provided by their habitats and the environment they live in. 
For instance, insects which are less vulnerable to radiation, were extinct
only in the greatest extinction - the end-Permian extinction $251~Myr$
ago.  Even then only 8 out of 27 orders were extinct compared with a
global species extinction that ranged between 80\% to 95\% (4). Also
plants which are less vulnerable to ionizing radiation suffered lower
level of extinction. Terrain, underground and underwater sheltering
against a complete extinction on land and in deep waters may explain why
certain families on land and in deep waters were not extinct even in the
great extinctions, while most of the species in shallow waters and on the
surface were extinct (4). Mountain shadowing, canyons, caves, underground
habitats, deep underwater habitats and high mobility may also explain why
many species like crocodiles, turtles, frogs,  (and most
freshwater vertebrates), snakes, deep sea organisms and birds were little
affected in the Cretaceous/Tertiary (K/T) boundary extinction which 
claimed the life of the big dinosaurs and pterosaurs. In particular,
fresh  underground waters in rivers and lakes are less polluted 
with radioisotopes and poisons produced by the CRB than sea
waters and may explain the survival of freshwater amphibians.  
 
{\bf Geological Signatures:}
The terrestrial deposition of the primary CRB nuclides  or the  
production of stable nuclides in the atmosphere or in the 
surface by the CRB is too small to be detectable. In particular, the 
proposed mechanism cannot 
explain the surface enrichment at the K/T boundary
by about $3\times 10^5~tons$ of iridium. Alvarez et al. (3)  
suggested that the impact of extraterrestrial asteroid, 
with an Ir abundance similar to that observed in early solar system 
Chondritic meteorites whose Ir abundance is larger than 
crustal Ir abundance by $\sim 10^4$, could cause the Ir 
anomaly and explain the mass extinction at the K/T boundary.
However, no significant Ir enrichment was found in all other mass
extinctions. Moreover, other isotopic anomalies due to meteoritic 
origin have not been found around the K/T boundary; in particular 
the As/Ir and Sb/Ir ratios are three orders of magnitude greater
than chondritic values but are in accord with a mantle origin (27). 
Extensive iridium measurements showed that the anomaly does not appear
as a single spike in the record, indicative of an instantaneous event, 
but rather occur over a measurable time interval of 10 to 100 kyr or
possibly longer (28). That and the high abundance of Ir in 
eruptive  magma led to the suggestion that the iridium
anomaly is due to a global volcanic activity over 10 to 100 kyr at the K/T
boundary (28,29) which also caused the K/T extinction:
Eruptions have a variety of short term effects,
including cooling from both dust and sulfates ejected into the stratosphere,
acid rain, wildfires, release of poisonous elements and increase 
in ultraviolet radiation from ozone-layer depletion. But, examination of
major volcanic eruptions in the past $100~ Myr$ have shown that none of 
them greatly affected the diversity of regional and global life on land
or in the oceans (4). 

A CRB will enhance the abundance of stable cosmogenic isotopes in the
geological layer corresponding to the CRB event, but, the enrichment may
be negligible compared to their accumulation through long terrestrial
exposure of the geological layers to galactic cosmic rays prior to the
CRB. However, CRB enrichment of sediments with unstable
radioisotopes of mean lifetimes much shorter than the age of the solar
system, $\tau\ll t_\odot\approx 4570 Myr$, but comparable to the
extinction times, may be detectable through low traces mass spectrometry.
In particular, fission of long lived
terrestrial nuclei, such as $^{238}$U and $^{232}$Th, by shower
particles, and capture of shower particles by such nuclei, may lead to
terrestrial production of, e.g.,  $^{129}$I with
$\tau=15~Myr$, $^{146}$Sm with $\tau= 146~Myr$, $^{205}$Pb with
$\tau=43~Myr$ and $^{244}$Pu with $\tau=118~Myr$, respectively. These
radioisotopes may have been buried in underwater sediments and
underground rocks which were protected from further exposure to cosmic
rays. The main background to such a CRB signature is the continuous
deposition by cosmic rays and by meteoritic impacts on land and sea.
Cosmic rays may include these trace radioisotopes due to nearby sources
(e.g., supernova explosions) and because of spallation of stable 
cosmic ray nuclei in collisions with interstellar gas. 
Meteorites may include these trace elements due to a long exposure in
space to cosmic rays. 

Finally, large enhancement of TeV cosmic ray tracks in  magma from
volcanic eruptions coincident with extinctions may 
also provide fingerprints for CRB extinctions. 

\section{Rate of Mass Extinctions}
Assuming that the spatial distribution of NS binaries in the MW follow the 
distribution of single pulsars (11),
\begin{equation} 
dN\propto e^{-R^2/2R_0^2}e^{-\vert z\vert /h}RdRdz,
\end{equation} 
with a disc scale length, $R_0\sim 4.8~kpc$, and a scale height, $h> 
0.5~kpc$
perpendicular to the disc and independent of disc position, we find that
the average rate of NS-NS mergers within $\sim 1~kpc$ from planet Earth is
$\sim 10^{-8}~yr^{-1}$. It is consistent with the 5 big  
extinctions which have occurred during the last $600~Myr$   
in the Paleozoic and Mesozoic eras. The relative strengths of these
extinctions may reflect mainly different distances from the CRBs.
Beyond $\sim 1~ kpc$ from the explosion
the galactic magnetic field begins to disperse the CRB and suppresses
its lethality.  Such CRBs, if not too far,  can still cause 
partial extinctions at a higher rate and induce biological mutations 
which may lead to the appearance of new species.

The galactic rate of SN explosions is $\sim 100~y^{-1}.$ The range of
debris from SN explosions in the interstellar medium is shorter than
$10~pc$. The rate of SN explosions within a distance of $10~pc$ from Earth
which follows from eq. 7 is $R_{SN}(<10~pc)\approx 10^{-10}y^{-1}$. High
energy cosmic rays which, perhaps, are produced in the SN remnant by shock
acceleration, carry only a small fraction of the total explosion energy
and arrive spread in time due to their diffusive propogation in the
interstellar magnetic field.  Also neutrino and light emissions similar to
those observed in SN1987A, at a distance of a few $pc$ cannot cause a mass
extinction. 
   
\section{Conclusions}
Cosmic Ray Bursts from neutron star mergers may have caused
the massive continental and marine life
extinctions which 
interrupted the diversification of life  on our planet. 
Their rate is consistent
with the observed rate of mass extinctions in the past $570~Myr.$ 
They may be able to explain the complicated  biological and geographical
extinction patterns. Biological mutations induced by the ionizing
radiations which are produced by the CRBs may explain
the appearance of completely new species after extinctions.
A first examination suggests a significant correlation
between the biological extinction pattern of different species and their
exposure and vulnerability to the ionizing radiation produced by a CRB.
The iridium enrichment around the Cretaceous/Tertiary extinction that
claimed the life of the dinosaurs and pterosaurs cannot be due to a CRB.
It may have been caused by intense volcanic eruptions around that
extinction. Isotopic anomaly signatures of CRB extinctions may be present  
in the geological layers which recorded the extinctions.   
Elaborate investigations of 
the effects of CRBs from relatively nearby neutron star mergers
and their biological, radiological and geological fingerprints  
are needed before reaching a firm conclusion
whether the massive extinctions during the long history of planet 
Earth  were caused by CRBs from neutron stars mergers. If nearby  
neutron star mergers are responsible for mass extinctions, then 
an early warning of future extinctions
due to neutron star mergers can be obtained by identifying, mapping and
timing all the nearby binary neutron stars systems. A final warning for an
approaching CRB from a nearby neutron-stars merger will be provided few
days before its arrival by a gamma ray burst produced by the approaching CRB.

\noindent
{\bf Acknowledgement}: This research was supported in part by the 
Technion  fund for promotion of research. 

\parindent 0cm
\centerline{References}

1. M.J. Benton, Science {\bf 278}, 52 (1995) and references 
therein. 

2. D.H. Erwin, Scientific American, {\bf 275}, 72 (1996) and 
references therein.

3. L.W. Alvarez et al., Science {\bf 208}, 1095 (1980).

4. D.H. Erwin, Nature, {\bf 367}, 231 (1994) and references therein.

5. S.E. Thorsett, ApJ. {\bf 444}, L53 (1995).

6. M.A. Ruderman, Science, {\bf 184}, 1079 (1974).  

7. R.A. Hulse \& J. Taylor, ApJ. {\bf 195}, L51 (1975).

8. G.H. Stokes et al., ApJ. {\bf 294}, L21 (1991).

9. A. Wolszcan, Nature {\bf 350}, 688 (1991). 

10. S.B. Anderson et al.,  Nature {\bf 346}, 42 (1990).

11. R.N. Manchester \& J.H. Taylor, {\it Pulsars}, (Freeman, San Francisco
1977); J.H. Taylor \& J.M. Weisberg, ApJ. {\bf 345}, 434 
(1989).

12. E.S. Phinney, ApJ. {\bf 380}, L17 (1992).

13. S.J. Curran, \& D.L. Lorimer, MNRAS {\bf 276}, 347 (1995).

14. See, e.g., S.L. Shapiro \& S.A. Teukolsky, {\it Black Holes, White
   Dwarfs \& Neutron Stars} (A. Wiley Intersc. Pub. 1983). 

15. G. Fishman, private communication.

16. H.T. Janka, \& M. Ruffert, Astr. \& Ap. {\bf 307},
L33 (1996).  

17. N.J. Shaviv, \& A. Dar,  Submitted to PRL, Astro-ph 
9606032. 
   
18. G.J. Fishman \& C.A.A. Meegan, Ann. Rev. Astr. 
 Ap. {\bf 33}, 415 (1995).  

19. B. Paczynski, AIP (Conf. Proc.) {\bf 280}, 981 (1993).

20. V.S. Berezinsky et al., {\it Astrophysics of Cosmic Rays},
(North Holland 1990) p. 71. 

21. C.C. Wu et al., ApJ. {\bf 416}, 247 (1993). 
 
22. A. Dar, A. Laor, \& N.J. Shaviv, to be published. 

23. J.W. Elbert, Proc. DUMAND Workshop (ed. A. Roberts) {\bf 2}, p. 101
(1978).

24. L. Montanet, et al., Phys. Rev. D {\bf 50},   
     1173-1826 (1994).
  
25. A. Dar, Phys. Rev. Lett. {\bf 51}, 227 (1983).

26. J.A.E. Stephenson et al., Nature {\bf 352}, 137 (1991).

27. C.B. Officer, \& C.L. Drake, Science, {\bf 227}, 1161 {1985}. 

28. See, e.g., C.B. Officer et al., Nature {\bf 326}, 143 (1987).

29. See, e.g., L. Zhao \& F.T. Kyt, Earth planet Sci. Lett. {\bf 90}, 
411 (1988).  

\end{document}